\documentclass[epj]{webofc}
\usepackage[varg]{txfonts}   
\usepackage{slashed}
\woctitle{MESON2014 - the 13$^\textrm{th}$ International Workshop on Meson Production, Properties and Interaction}
\begin{document}
\selectlanguage{english}
\title{%
	Heavy-quark expansion for D and B mesons in nuclear matter
}

\author{Thomas Buchheim\inst{1}\fnsep\thanks{\email{t.buchheim@hzdr.de}} \and
        Thomas Hilger\inst{2} \and
        Burkhard K\"{a}mpfer\inst{1,3} 
}

\institute{Helmholtz-Zentrum Dresden-Rossendorf, PF 510119, D-01314 Dresden, Germany 
\and
           University of Graz, Institute of Physics, NAWI Graz, A-8010 Graz, Austria
\and
           Technische Universit\"{a}t Dresden, Institut f\"{u}r Theoretische Physik, D-01062 Dresden, Germany
          }

\abstract{%
The planned experiments at FAIR enable the study of medium modifications of \(D\) and \(B\) mesons in (dense) nuclear matter. Evaluating QCD sum rules as a theoretical prerequisite for such investigations encounters heavy-light four-quark condensates. We utilize an extended heavy-quark expansion to cope with the condensation of heavy quarks.
}
\maketitle
\section{Introduction}
\label{sec:intro}

The forthcoming experimental perspectives for in-medium heavy-light quark meson spectroscopy, in particular at FAIR, are accompanied by the need for sophisticated theoretical analyses, e.\,g.\ \cite{tolos09,blaschke,yasuisudoh13a,he,hilger09,wang11}. When utilizing QCD sum rules \cite{svz79,rry,narison}, this requires a thorough discussion of heavy-quark condensates in general, and, in particular, in the nuclear medium \cite{suzukigubler13}. Therefore, the heavy-quark expansion (HQE), originally developed for the heavy two-quark condensate \(\langle \bar{Q}Q \rangle\) in vacuum, is extended to four-quark condensates and to the in-medium case, thus going beyond previous approaches, e.\,g.\ \cite{narison13}. Specific formulas are derived and presented.

\section{Recollection: HQE in vacuum}
\label{sec:HQE_vac}
 
In \cite{genbroad}, a general method is introduced for vacuum condensates involving heavy quarks \(Q\) with mass \(m_Q\). The heavy-quark condensate is considered as the one-point function
\begin{align}
	\langle 0 | \bar{Q}Q | 0 \rangle = -i \int \frac{d^4p}{(2\pi)^4} \langle 0 | \text{Tr}_\text{c,D}\; S_{\!\! Q}(p) | 0 \rangle
	\label{eq:formula_HQEofQQ}
\end{align}
expressed by the heavy-quark propagator \(S_{\!\! Q}\) in a weak classical gluonic background field in Fock-Schwinger gauge, \(S_{\!\! Q}(p) = \sum_{k=0}^\infty S^{(k)}_{\!\! Q} (p)\) with \(S^{(k)}_{\!\! Q} (p) = (-1)^k S_{\!\! Q}^{(0)}(p) \gamma^{\mu_1}\tilde{A}_{\mu_1} S_{\!\! Q}^{(0)}(p) \ldots \gamma^{\mu_k}\tilde{A}_{\mu_k} S_{\!\! Q}^{(0)}(p)\), incorporating the free heavy-quark propagator \(S_{\!\! Q}^{(0)}(p) = (\gamma^\mu p_\mu + m_Q)/(p^2 - m_Q^2)\) and the derivative operator \(\tilde{A}\) emerging from a Fourier transform defined as \(\tilde{A}_\mu = \sum_{m=0}^\infty \tilde{A}_\mu^{(m)}\) with \(\tilde{A}_\mu^{(m)} = -\frac{(-i)^{m+1} g}{m!(m+2)} \left( D_{\alpha_1} \ldots D_{\alpha_m} G_{\mu\nu}(x) \right)_{x=0} \partial^\nu \partial^{\alpha_1} \ldots \partial^{\alpha_m}\) \cite{hilger11,zsch11}. In this way, the heavy-quark propagator interacts with the complex QCD ground state via soft gluons generating a series expansion in the inverse heavy-quark mass. The compact notation \eqref{eq:formula_HQEofQQ} differs from \cite{genbroad}, but provides a comprehensive scheme easily extendable to in-medium condensates. The first HQE terms of the heavy two-quark condensate \eqref{eq:formula_HQEofQQ} reproduce \cite{genbroad}:
\begin{align}
	\langle 0 | \bar{Q}Q | 0 \rangle & = - \frac{g^2}{48\pi^2 m_Q} \langle G^2 \rangle - \frac{g^3}{1440\pi^2 m_Q^3} \langle G^3 \rangle - \frac{g^4}{120\pi^2 m_Q^3} \langle (DG)^2 \rangle + \ldots 
	\label{eq:HQEofQQ}\\
										& = \!\!\!\!\! \parbox[c][15mm][c]{20mm}{\mbox{\includegraphics[width=2cm]{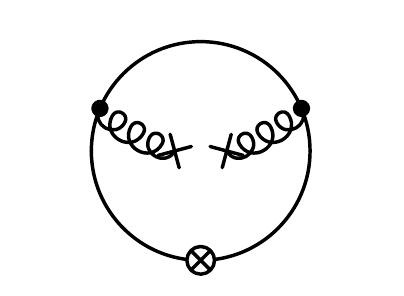}}} \!\!\!\!\!
											+ \; \left( \!\!\!\!\! \parbox[c][15mm][c]{20mm}{\mbox{\includegraphics[width=2cm]{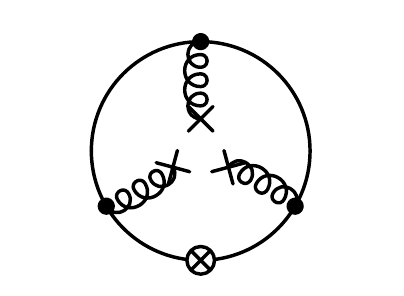}}} \!\!\!\!\!
											          + \!\!\!\!\! \parbox[c][15mm][c]{20mm}{\mbox{\includegraphics[width=2cm]{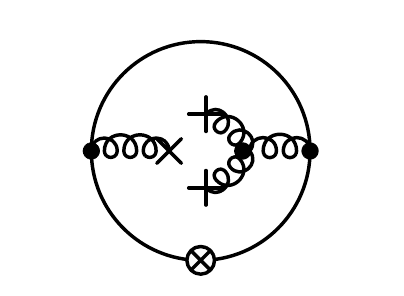}}} \!\!\!\!\! \right) \;
											+ \!\!\!\!\! \parbox[c][15mm][c]{20mm}{\mbox{\includegraphics[width=2cm]{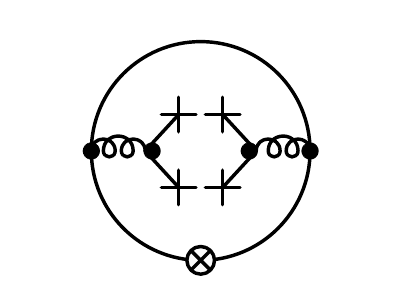}}}\!\!\!\!\! + \; \ldots \nonumber
\end{align}
with the notation
\begin{align}
	\langle G^2 \rangle & = \langle 0 | G^A_{\mu\nu} G^{A\,\mu\nu} | 0 \rangle \, , \\[1ex]
	\langle G^3 \rangle & = \langle 0 | f^{ABC} G^A_{\mu\nu} {G^{B\,\nu}}_\lambda G^{C\,\lambda\mu} | 0 \rangle \, , \\
	\langle (DG)^2 \rangle & = \langle 0 | \bigg( \sum_{f} \bar{q}_{f} \gamma_\mu t^A q_{f} \bigg)^2 | 0 \rangle \, .
\end{align}
The graphic interpretation of the terms in \eqref{eq:HQEofQQ} is depicted too: the solid lines denote the free heavy-quark propagators and the curly lines are for soft gluons whose condensation is symbolized by the crosses, whereas the heavy quark-condensate is symbolized by the crossed circles \cite{bagan86}.  An analogous expression for the mixed heavy-quark gluon condensate can be obtained along those lines which contains, however, a term proportional to \(m_Q\). The leading-order term in \eqref{eq:HQEofQQ} was employed already in \cite{svz79} in evaluating the sum rule for charmonia.

The vacuum HQE method was rendered free of UV divergent results for higher mass-dimension heavy-quark condensates by requiring at least one condensing gluon per condensed heavy-quark \cite{bagan85,bagan86}, which prevents unphysical results, where the condensation probability of heavy-quark condensates rises for an increasing heavy-quark mass.

\section{Application of HQE to in-medium heavy-light four-quark condensates}
\label{sec:HQE_qqQQ}

The above method can be extended to in-medium situations. Our approach contains two new aspects: (\(i\)) formulas analogous to equation~\eqref{eq:formula_HQEofQQ} are to be derived for heavy-quark condensates, e.\,g.\ \(\langle \bar{Q} \slashed{v} Q \rangle\), \(\langle \bar{Q} \slashed{v} \sigma G Q \rangle\), \(\langle \bar{q} \slashed{v} t^A q \bar{Q} \slashed{v} t^A Q \rangle\), which additionally contribute to the in-medium operator product expansion (OPE) and (\(ii\)) medium-specific gluonic condensates, e.\,g.\ \(\langle G^2/4 - (vG)^2/v^2 \rangle\), \(\langle G^3/4 -  f^{ABC} G^A_{\mu\nu} {G^{B\,\nu}}_\lambda G^{C\,\lambda\kappa} v^\mu v_\kappa / v^2 \rangle\), enter the HQE of heavy-quark condensates for both, vacuum and additional medium condensates, where \(\langle \ldots \rangle\) denotes Gibbs averaging. 

We are especially interested in heavy-light four-quark condensates entering the OPE of \(D\) and \(B\) mesons, inter alia, in terms corresponding to the next-to-leading-order perturbative diagrams with one light-quark (\(q\)) and one heavy-quark (\(Q\)) line cut. There are 24 two-flavour four-quark condensates in the nuclear medium \cite{thomas07} represented here in a compact notation by \(\langle \bar{q} \Gamma T^A q \bar{Q} \Gamma' T^A Q  \rangle\), where \(\Gamma\) and \(\Gamma'\) denote Dirac structures and \(T^A\) with \(A=0,\ldots,8\) are the generators of \(S\! U(3)\) supplemented by the unit element (\(A=0\)). We obtain the analogous formula to \eqref{eq:formula_HQEofQQ} for heavy-light four-quark condensates:
\begin{align}
	\langle \bar{q} \Gamma T^A q \bar{Q} \Gamma' T^A Q  \rangle = -i \int \frac{d^4p}{(2\pi)^4} \langle \bar{q} \Gamma T^A q \; \text{Tr}_\text{c,D} \left[\Gamma' T^A S_{\!\! Q}(p) \right] \rangle \, .
\end{align}
The leading-order terms of this HQE are obtained for the heavy-quark propagators \(S_{\!\! Q}^{(1)}\) containing \(\tilde{A}_\mu^{(1)}\) and \(S_{\!\! Q}^{(2)}\) with leading-order background fields \(\tilde{A}_\mu^{(0)}\):
\begin{align}
	\langle \bar{q} \Gamma T^A q \bar{Q} \Gamma' T^A Q  \rangle & = -i \int \frac{d^4p}{(2\pi)^4} \langle \bar{q} \Gamma T^A q \; \text{Tr}_\text{c,D} \left[\Gamma' T^A \left( S_{\!\! Q}^{(1)}(p) + S_{\!\! Q}^{(2)}(p) + \ldots \right) \right] \rangle 
	\label{eq:expan_HQEofQQqq}\\[2ex]
	& = \langle \bar{q} \Gamma T^A q \bar{Q} \Gamma' T^A Q  \rangle^{(0)} + \langle \bar{q} \Gamma T^A q \bar{Q} \Gamma' T^A Q  \rangle^{(1)} + \ldots 
	\label{eq:firstterms_HQEofQQqq}\\
	& = \!\!\!\!\! \parbox[c][15mm][c]{20mm}{\mbox{\includegraphics[width=2cm]{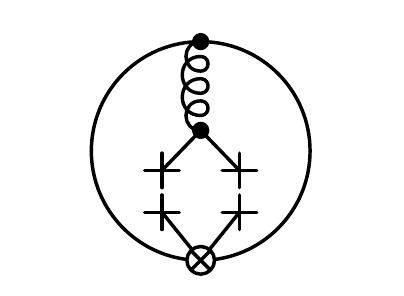}}} \!\!\!\!\!
		+ \!\!\!\!\! \parbox[c][15mm][c]{20mm}{\mbox{\includegraphics[width=2cm]{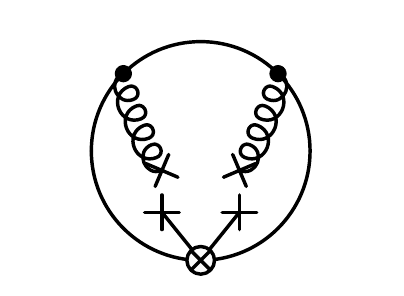}}}\!\!\!\!\! + \; \ldots \, . \nonumber
\end{align}
Evaluation of the first term of the expansion \eqref{eq:firstterms_HQEofQQqq} for the complete list of two-flavour four-quark condensates in \cite{thomas07} gives three non-zero results:
\begin{align}
	\langle \bar{q} \gamma^\nu t^A q \bar{Q} \gamma_\nu t^A Q  \rangle^{(0)} & = -\frac{2}{3}\frac{g^2}{(4\pi)^2}\left( \log\frac{\mu^2}{m_Q^2} + \frac{1}{2} \right) \langle \bar{q} \gamma^\nu t^A q \sum_f \bar{q}_f \gamma_\nu t^A q_f  \rangle \, , \\
	\langle \bar{q} \slashed{v} t^A q \bar{Q} \slashed{v} t^A Q  \rangle^{(0)} & = -\frac{2}{3}\frac{g^2}{(4\pi)^2}\left( \log\frac{\mu^2}{m_Q^2} + \frac{2}{3} \right) \langle \bar{q} \slashed{v} t^A q \sum_f \bar{q}_f \slashed{v} t^A q_f  \rangle \, , \\
	\langle \bar{q} t^A q \bar{Q} \slashed{v} t^A Q  \rangle^{(0)} & = -\frac{4}{3}\frac{g^2}{(4\pi)^2}\left( \log\frac{\mu^2}{m_Q^2} - \frac{1}{8} \right) \langle \bar{q} t^A q \sum_f \bar{q}_f \slashed{v} t^A q_f  \rangle \, ,
\end{align}
where logarithmic singularities are calculated in the \(\overline{\text{MS}}\) scheme, \(\mu\) is the renormalization scale, and \(t^A= T^A\) for \(A = 1,\ldots,8\). The non-zero contributions for the second term of \eqref{eq:firstterms_HQEofQQqq} read
\begin{align}
	\langle \bar{q}q \bar{Q}Q  \rangle^{(1)} & = -\frac{1}{3}\frac{g^2}{(4\pi)^2} \frac{1}{m_Q} \langle \bar{q}q G^A_{\mu\nu} G^{A\, \mu\nu} \rangle \, , \\
	\langle \bar{q} t^A q \bar{Q} t^A Q  \rangle^{(1)} & = -\frac{1}{6}\frac{g^2}{(4\pi)^2} \frac{1}{m_Q} \langle d^{ABC} \bar{q} t^A q G^B_{\mu\nu} G^{C\, \mu\nu} \rangle \, , \\
	\langle \bar{q} \gamma_5 q \bar{Q} \gamma_5 Q  \rangle^{(1)} & = - \frac{1}{4}\frac{g^2}{(4\pi)^2} \frac{1}{m_Q} \langle i \bar{q} \gamma_5 q G^A_{\mu\nu} G^{A}_{\alpha\beta} \varepsilon^{\mu\nu\alpha\beta} \rangle \, , \\
	\langle \bar{q} \gamma_5 t^A q \bar{Q} \gamma_5 t^A Q  \rangle^{(1)} & = - \frac{1}{8}\frac{g^2}{(4\pi)^2} \frac{1}{m_Q} \langle i d^{ABC} \bar{q} \gamma_5 t^A q G^B_{\mu\nu} G^{C}_{\alpha\beta} \varepsilon^{\mu\nu\alpha\beta} \rangle \, , \\
	\langle \bar{q} \slashed{v} q \bar{Q}Q  \rangle^{(1)} & = -\frac{1}{3}\frac{g^2}{(4\pi)^2} \frac{1}{m_Q} \langle \bar{q} \slashed{v} q G^A_{\mu\nu} G^{A\, \mu\nu} \rangle \, , \\
	\langle \bar{q} \slashed{v} t^A q \bar{Q} t^A Q  \rangle^{(1)} & = -\frac{1}{6}\frac{g^2}{(4\pi)^2} \frac{1}{m_Q} \langle d^{ABC} \bar{q} \slashed{v} t^A q G^B_{\mu\nu} G^{C\, \mu\nu} \rangle \, , \\
	\langle \bar{q} \sigma_{\mu\nu} t^A q \bar{Q} \sigma^{\mu\nu} t^A Q  \rangle^{(1)} & = -\frac{5}{6}\frac{g^2}{(4\pi)^2} \frac{1}{m_Q} \langle f^{ABC} \bar{q} \sigma_{\mu\nu} t^A q G^{B\,\nu}{}_\lambda G^{C\, \lambda\mu} \rangle \, , \\
	\langle \bar{q} \sigma_{\mu\nu} t^A q \bar{Q} \sigma_{\alpha\beta} t^A Q g^{\mu\alpha} v^\nu v^\beta \rangle^{(1)} & = - \frac{5}{3}\frac{g^2}{(4\pi)^2} \frac{1}{m_Q} \langle f^{ABC} \bar{q} \sigma_{\mu\nu} t^A q G^{B\,\nu}{}_\alpha G^{C\, \alpha\beta} v^\mu v_\beta \rangle \, , \\
	\langle \bar{q} \gamma_5 \gamma_\lambda t^A q \bar{Q} \sigma_{\mu\nu} t^A Q \varepsilon^{\mu\nu\lambda\tau} v_\tau \rangle^{(1)} & = -\frac{5}{6}\frac{g^2}{(4\pi)^2} \frac{1}{m_Q} \langle f^{ABC} \bar{q} \gamma_5 \gamma_\lambda t^A q G^B_{\alpha\beta} G^{C\, \beta}{}_\gamma \varepsilon^{\gamma\alpha\lambda\tau} v_\tau \rangle \, ,
\end{align}
where \(f^{ABC}\) is the anti-symmetric structure constant of the color group and the corresponding symmetric object \(d^{ABC}\) is defined by the anti-commutator \(\{ t^A , t^B \} = \delta^{AB}/4 + d^{ABC} t^C\).

\section{Summary and Conclusions}
\label{sec:sum}

The extension of the OPE for QCD sum rules of \(\bar{q}Q\) and \(\bar{Q}q\) mesons by four-quark condensates to mass dimension 6 yields heavy-light condensate contributions requiring HQE in a nuclear medium. The necessary steps to generalize the vacuum HQE \cite{genbroad} to cover in-medium situations are described and a general formula for the HQE of in-medium heavy-light four-quark condensates is presented. The two leading-order terms of this expansion for the complete list of two-flavour four-quark condensates \cite{thomas07} have been evaluated. In leading-order the results contain known condensate structures, thus, reducing the number of condensates entering the sum rule evaluation of mesons composed of a heavy and a light quark. It can be seen that the series does not exhibit a simple expansion in \(1/m_Q\), not even in vacuum. Therefore, the lowest order terms are not suppressed by inverse powers of \(m_Q\) as for \(\langle \bar{Q}Q \rangle\), challenging the omission of heavy-light four-quark condensates, as often done in previous sum rule analyses.

%
%

\begin{acknowledgement}
This work is supported by BMBF 05P12CRGHE and the Austrian Science Fund (FWF) under project no. P25121-N27.
\end{acknowledgement}
%
\bibliography{lit}
%
%
%
%

\end{document}